\documentclass[12pt]{article}
\usepackage[dvips]{graphics}

\def\be{\begin{equation}}
\def\ee{\end{equation}}
\def\bea{\begin{eqnarray}}
\def\eea{\end{eqnarray}}

\def\m{\m_\nu}
\def\la{\langle}
\def\ra{\rangle}
\def\gd{$G^{(\rm D)}$ }
\newtheorem{them}{Theorem}

\newtheorem{lemma}[them]{Lemma}

\author{C. King$^1$ and F. Y. Wu$^2$  \\
  \\
$^1$Department of Mathematics\\
$^2$Department of Physics\\
  Northeastern University \\
 Boston,
Massachusetts 02115}

\title{New correlation duality relations for the planar Potts model}

\begin{document}
\maketitle

\bigskip
\begin{abstract}
We introduce  a new method to generate duality relations for
correlation functions of the Potts model on  planar graphs.
The method extends previously known results, by allowing
the consideration of the correlation function for arbitrarily placed vertices
 on the graph.  We show that  generally
it is linear combinations of correlation functions,
not the individual correlations,
that are related by dualities.
 The method is illustrated in several non-trivial cases,
and the relation to earlier results is  explained.
A graph-theoretical formulation of our results in terms of  rooted
dichromatic, or Tutte, polynomials is also given.
\end{abstract}
\vskip 1cm
\noindent

\newpage

\section{Introduction}
The Potts model,   a generalization of the Isling model
to $q$ spin components introduced
 by Potts \cite{potts} in 1952, has been at the forefront of research interest
for almost five decades.    While the critical properties
of planar models are now largely known \cite{baxter} - \cite{wureview1},
 much less
is known about their correlation functions.

A key element of planar spin models is the notion of duality relations.
 Every planar  graph, or lattice, $G$ has an associated
dual planar graph $G^{(\rm D)}$, and it is well-known \cite{potts,wangwu}
that the
partition functions of the Potts model on $G$  and \gd are proportional to
each other when their couplings are correctly  related.
  In the case of  $q=2$, the Ising model, it is  also known \cite{watson}
that a duality
relation exists between certain two-spin
correlation functions.
 It has long been suspected that the  correlations of the Potts models
on $G$ and $G^{(\rm D)}$ should be similarly related.
Indeed, in a series of recent papers we have obtained correlation dualities for
vertices (spins)  residing on the boundary of one face
\cite{wu} - \cite{WuLu} or two faces \cite{Ki} of $G$,
and reformulated the 1--face results in graph-theoretical terms
\cite{wukinglu}.
   In this paper we consider the $N$-face problem,
namely, correlation duality relations for vertices residing on $N$
faces of $G$ for general $N$.
As we shall see, a new picture emerges  for  $N\geq 2$.
   Instead of duality relations
between individual correlation functions,
 we now obtain
relations between linear combinations of
correlation functions.

The organization of this paper is as follows.
In Sec. 2 we establish notations and briefly review
the  Fortuin-Kasteleyn (F-K) representation of the Potts model, and
derive
the duality relation for the partition function.
Using  two identities which we establish
as  lemmas, we introduce in Section 3 a method
which allows us to generate correlation duality relations.  While
our method does not necessarily generate all such duality relations,
we show  that generally
the duality relates linear combinations of, rather
than individual, correlation functions.
Explicit examples are given in Section 4 and the relation to earlier
results is explained.  In Section 5 we reformulate our results
in graph-theoretical terms as rooted dichromatic, or Tutte, polynomials.

\section{The setup}
\subsection{The Fortuin-Kasteleyn representation}
Consider a finite, planar connected graph $G$ with  vertex set $V$,
edge set $E$,
and face set $F$. We write $|V|$ for the size of the set $V$
 and similarly for $E$,  $F$, and other terms. Let
$\sigma$ denote an assignment of  colors, or states,
 to the vertices of $G$, which is
a map from
$V$ to the color set
${\cal C} = \{1,\dots,q\}$.
Then the partition function of the  Potts
model on $G$ is
\be\label{def:Z}
Z = \sum_{\sigma} \,\prod_{\la i,j \ra} \,
e^{ K \delta({\sigma}_i,
{\sigma}_j)}
\ee
where the summation is over
the $|V|^q$ coloring assignments $\sigma$,  the product runs
over all nearest neighbor spin pairs interacting with the coupling
parameter $K$,
$\delta$  is the Kronecker delta function, and $\sigma_i$ is the
color of the $i$-th vertex.

The Fortuin-Kasteleyn (F-K) representation \cite{fk} of the Potts model
replaces the configurational sum  in (\ref{def:Z}) by a sum over edge sets.
Let $S \subseteq E$ be an edge set on the graph $G$.
If we remove from $G$ all the edges which are {\it not} in $S$, we
obtain a subgraph which contains all the vertices, and which
in general has many connected components. This is called a spanning
subgraph of $G$.  We write $|S|$ to denote the number of edges in $S$,
and  write $p(S)$ to denote the number of connected components
in the spanning subgraph defined by $S$. In this picture, any isolated vertex
in the spanning subgraph counts as a connected component, and so
contributes a summand 1 in the counting of $p(S)$.
  Define
\be\label{def:v}
v = e^K - 1
\ee
so that we have
\be\label{fk}
e^{ K \delta({\sigma}_i,\sigma_j)} = 1+v\, \delta({\sigma}_i,\sigma_j)
\ee
The substitution of (\ref{fk}) into (\ref{def:Z}) now gives rise  to
  the F-K representation of the partition function
\be\label{Z-F-K}
Z(G;q,v)\equiv Z = \sum_{S \subseteq E} v^{|S|} \,\,q^{p(S)}\label{part}
\ee
The number of colors $q$  appears as a variable in
(\ref{Z-F-K}), rather than  a summation
limit as in  (\ref{def:Z}). For this reason
the F-K representation (\ref{part}) forms the basis of essentially
all  analyses of the Potts model.

\subsection{Dual partition functions}
As a precursor to discussions  in later
sections, we give here a graph-theoretical derivation of  the well-known
duality relation
for the partition function.

The standard dual graph of $G$ is constructed by placing vertices
in the faces of $G$, and connecting them with edges which are transverse to the
edges of $G$.  A pair of transverse edges are said to be
dual to each other.
 The resulting dual  graph $G^{(\rm D)}$ has  $|F|$ vertices,
$|E|$ edges and $|V|$ faces, and is also planar.
  For a given edge set $S$ on $G$,
we define its dual edge set $S^{(\rm D)}$ on \gd
by requiring that an edge of \gd is in $S^{(\rm D)}$ if and
only if its dual edge is not in $S$.  Thus,
\be\label{edgenumber}
|S|+|S^{(\rm D)}|=|E|
\ee
Moreover,  if we start  from $|V|$ isolated vertices
of $G$, and add one by one the edges of $S$, then each
edge reduces $p(S)$ by 1 unless it forms an  independent
cycle, in which case $p(S)$ is unchanged.  Therefore
 we have the relation
\bea\label{graphdual}
p(S) &=& |V|-|S| +c(S) \nonumber \\
    &=& |V|-|S| +p(S^{(\rm D)})-1
\eea
where  $c(S)$ is the number of independent cycles in the spanning subgraph
of $S$,
and is related to $p(S^{(\rm D)})$, the number of components  in $S^{(\rm D)}$,
 through the topological relation $c(S)=p(S^{(\rm D)})-1$.
Using (\ref{edgenumber})
and (\ref{graphdual})  we have  the
identity
\bea\label{iden1}
v^{|S|}\,q^{p(S)}& =& v^{|E|- |S^{(\rm D)}| } q^{|V|-|S| +p(S^{(\rm
D)})-1}\nonumber \\
                 & = & k\, (q/v)^{|S^{(\rm D)}|}\,q^{p(S^{(\rm D)})}
\eea
where
\be\label{k}
k=q^{1-|F|} v^{|E|}
\ee
and we have used the Euler relation
\be\label{euler}
|V|+|F| =|E| +2
\ee
Substituting (\ref{iden1}) into (\ref{part}), we see
that the Potts model partition functions on $G$ and $G^{(\rm D)}$  are
proportional,
when the couplings $K$ and  $K^*$ are related
 by the equation
\be\label{def:K^*}
vv^* = q
\ee
where $v^* = e^{K^*} - 1$.
The partition function duality is the statement that
\be\label{origdual}
 Z(G;q,v) =k \, Z(G^{(\rm D)};q,v^*)
\ee
where we have   made use of the  one-to-one correspondence
between edge  sets $S$ and  dual edge sets $S^{(\rm D)}$.
Note that we have
\be\label{check}
k\cdot k^{(\rm D)}= q^{1-|F|} v^{|E|} \cdot q^{1-|V|} (v^*)^{|E|} =1
\ee
 which says  that the dual of the dual
partition function is the partition function itself.

\subsection{The correlation function}
The  Potts correlation function is defined
as the probability that a given set of vertices are assigned
fixed colors \cite{wureview}.
 Specifically, let $R \subseteq V$ be any subset of the vertices which are
assigned
fixed colors. We shall call these the
{\it roots} for the correlation functions, and the graph with roots a {\it
rooted} graph.
Let the $|R|$ roots lie on the boundaries of $N$ distinct faces of $G$,
which  will be
called the {\it external} faces of the graph.
For simplicity, we shall assume that external faces are not
adjacent to each other, namely they do not share any vertices or
boundary edges.

Let $c: R
\rightarrow {\cal C}$ be any assignment of colors on the roots, so that
$c_i$ is the color assigned to the root $i$, and   we write
\be
{\delta}_{c}(\sigma) = \prod_{i \in R}
{\delta}(\sigma_i, {c}_i)  \label{delta}
 \ee
Then the correlation function for the assignment $c$ on the root set $R$ is
defined
to be the expectation value of $\delta_c(\sigma)$, namely,
\be\label{def:corr}
P_c \equiv
\la {\delta}_{c}(\sigma) \ra = Z(G;q,v)^{-1}
{\sum_{\sigma}}\, {\delta}_{c}(\sigma)\prod_{\la i,j \ra} \, e^{ K
\delta({\sigma}_i,\sigma_j)}
 \ee
 where the summation in the right-hand side is
a {\it partial} partition function.
   Clearly, the explicit expression of $P_c$  depends on
the location of the roots as well as
the explicit color assignment $c$.

Every color assignment $c$ defines a partition $X$ of the roots, namely a
division of the $|R|$ roots into disjoint blocks, by the rule that two
vertices $i,j$ in
$R$ belong to the same block of $X$ if and only if $c_i = c_j$.
By the
symmetry of the Potts interactions, the correlation
(\ref{def:corr}) depends only on this partition defined by $c$,
not on the specific colors. We can therefore associate each partial
partition function to  a partition $X$.
Alternately,
for any partition $X$ of the roots, we have a  {\it partial} partition
function
\be\label{def:Z_X}
Z_X \equiv Z_{X}(G;q,v)
= {\sum_{\sigma}} \, {\delta}_{c}(\sigma)  \prod_{\la i,j \ra} \, e^{ K
\delta({\sigma}_i,\sigma_j)}
\ee
where $c$ is any color assignment  that produces the partition $X$.
Since $P_c$ and $Z_X$ are proportional,
 it is sufficient to discuss duality relations for  the partial partition
function $Z_X$.

The F-K representation also extends to the partial partition functions.
 For any edge set $S$, we   define a ``connected"
partition $\pi(S)$ of $R$ by the rule that
two roots $i,j$ belong to the same block if and only if they belong to the
same {\it connected} component of the spanning subgraph defined by $S$.
Using this notion, we can
separate the edge sets into classes
\be\label{def:calS1}
{\cal S}(X) = \{S \subseteq E \,:\, \pi(S) = X \}
\ee
 labeled by their corresponding
connected partitions of $R$.
 Clearly every edge set $S$ belongs to exactly one of these classes
${\cal S}(X)$.

Now the role of the factor $\delta_c(\sigma)$ in (\ref{def:corr})
and (\ref{def:Z_X}) is to restrict the   summations to
fixed colors for vertices in $R$.  The net result is that  each cluster
containing
roots contributes  a factor 1, instead of the  factor $q$
 in (\ref{part}).    This leads us to define, for any
partition $X$, the summation over edge sets $S$ with $\pi(S)=X$,
\be\label{def:F_X}
F_{X}(G;q,v) = \sum_{S \in {\cal S}(X)} v^{|S|} \,\, q^{p(S) - |X|},
\ee
where $|X|$ is the number of
blocks in
$X$.  The set of partitions of $R$ is partially ordered.   We write
$Y \preceq X$ to mean that every block in $Y$ is wholly contained in a
block in $X$ or, equivalently, $Y$ is a refinement of $X$.
Then we have the identity
 \be\label{F-Kcorr}
Z_X(G;q,v) = \sum_{Y \preceq X} F_Y(G;q,v)
\ee

\subsection{M\"obius inversion and planar identities}
It is well-known \cite{lintw} that the partially ordered sum
(\ref{F-Kcorr}) can be inverted, allowing us to write
$F_X$ as a linear combination of $Z_Y$. For $Y \preceq X$ define
the M\"obius inversion coefficient
\be
\mu (Y,X) = (-1)^{|Y| - |X|} \prod_{b\in X} \label{mu}
(n_{b}(Y) - 1)!
\ee
where the product on the right side runs over blocks $b$ in $X$, and
$n_{b}(Y)$ is the number of blocks of $Y$ that are contained in $b$.
Then we have for every partition $X$,
\be\label{Mobius}
F_X(G;q,v) = \sum_{Y \preceq X} \mu (Y,X) Z_Y(G;q,v)
\ee

One peculiarity for planar graphs is that
$F_X$ can be zero for certain partitions $X$. This phenomenon was explored
in detail in \cite{WuLu} in the case of a single external face.
In this case, one has $F_X =0$ if the  partition $X$ is non-planar (also
called a
crossing partition in the mathematics literature \cite{Stan}).
It then follows from (\ref{Mobius}) that whenever $X$ is a
non-planar partition, the partial partition functions
$Z_Y$ must satisfy the identity
\be
\sum_{Y \preceq X} \mu (Y,X) Z_Y(G;q,v) = 0 \label{Mobius1}
\ee
giving rise to
  ``sum-rule" relations for the partial partition functions \cite{wuhuang}.
In general, when there are several external faces,  identities of this
type also hold for any root set
$R$, although it seems to be a hard problem to decide if a given
partition is planar or not.

\section{Correlation duality relation}
\subsection{Dual rooted graph $G^*$}
In Section 2.2 we introduced the dual graph $G^{(\rm D)}$,
and described the well-known duality for partition functions
on $G$ and $G^{(\rm D)}$. In this section we introduce the
dual rooted graph $G^*$ which plays the corresponding role
for correlation functions.

Starting from a rooted graph with $|R|$ roots on $N$ external
faces,  one can construct a dual rooted graph $G^*$.
 The procedure of constructing $G^*$ for $N=1$ has been described in detail
 in  \cite{WuLu,wukinglu}.  The construction of $G^*$ for general $N$
 repeats the $N=1$ process for each external face, a prescription
we now briefly describe.


 Place $N$ extra vertices, one each in the center of each of the $N$ external
faces.  For each external face,
let $f$ be the extra vertex  and connect it to each root
on the boundary of this face by an edge. This gives a new graph $G''$
which has $N$ more vertices than $G$ and $|R|$ additional edges.
  The dual graph of $G''$
is also planar, and it has $N$ faces containing the $N$ extra vertices $f$.
Now remove all edges of these $N$ faces,  and the resulting graph is the
dual rooted graph $G^*$.

The vertices of $G^*$ residing inside the $N$ external
faces of $G$ are now the dual roots
and we denote this set by  $R^*$.  Clearly,
 the number of edges of $G^*$ is   $|E|$, and
we have $|R^*| =|R|$ (because external
faces are non-adjacent).  It is also clear that the total number
of vertices of $G^*$ is
\be\label{Vstar}
|V^*| = |F| + |R| -N
\ee

We can recover the standard dual graph $G^{(\rm D)}$ from $G^*$
by   {\it fusing}
 the dual roots $R^*$ inside each external face into  a single
dual vertex. This gives a one-to-one correspondence between edges
on $G^{(\rm D)}$ and $G^*$, and hence we {\it define} the dual
edge set $S^*$ on $G^*$ to be the same edge set
as $S^{(\rm D)}$ on $G^{(\rm D)}$.
Therefore in particular we have
\be
|S^*|=|S^{(\rm D)}|
\ee
Let $G^{(\rm F)}$ denote the graph generated by fusing all roots $R$
in $G$,
and $( {G^*})^{(\rm F)}$ the graph generated by fusing all roots $R^*$
in $G^*$.
Then, we have the identities
\be
( {G^*})^{(\rm F)} = G^{(\rm D)}\hskip .5cm
{\rm and} \hskip .5cm G ^{(\rm F)}=(G^*)^{(\rm D)}
\ee

\subsection{A preliminary relation}
Our results will depend crucially on the topology of the connectedness of the
partitions $\pi(S)$ and $\pi(S^*)$.
Fix an edge set $S$ on $G$, and let $X = \pi(S)$,
$Y= \pi(S^*)$.  Recall that $|X|$ is the number of blocks in $X$.
Also, let $n(X)$ be the number of clusters
of external faces on $G$ connected by $S$ (where we define two
external faces to be in the same cluster if a connected component
of $S$ contains roots on both faces). Similar
definitions apply to $|Y|$ and $n(Y)$.

Now $S^*$ and $S^{(\rm D)}$ are the same edge set,
and $G^{(\rm D)}$ is obtained from  $G^*$
 by fusing.
 Hence we have
\bea
p(S^*) &=& |Y| + \Delta \nonumber \\
p(S^{(\rm D)}) &=& n(Y) + \Delta
\eea
where $\Delta$ is the number of connected components in
the spanning graph generated by $S^*$  that
are {\it not} connected to any dual roots.
It follows that we have
\be\label{y}
|Y| - n(Y) = p(S^*) - p(S^{(\rm D)})
\ee
and similarly
\be\label{x}
|X| - n(X) = p(S) - p({S^*}^{(\rm D)})
\ee
where ${S^*}^{(\rm D)}$ on ${G^*}^{(\rm D)}$ is the edge set dual
to $S^*$ on $G^*$.
Combining (\ref{y}) and (\ref{x}) and making use of (\ref{graphdual}),
we obtain the identity
\bea\label{xy}
|X| +|Y|  - n(X) -n(Y)
                       &=& [p(S) - p(S^{(\rm D)})]+[ p(S^*) -p({S^*}^{(\rm
D)})] \nonumber \\
                        &=& [|V|-|S|-1] + [|V^*|-|S^*|-1] \nonumber \\
                       &=& |V^*| - |F| \nonumber \\
                       &=& |R| - N
\eea
where use has also been made of (\ref{Vstar}), the Euler relation
(\ref{euler}),
and  the relation
\be\label{edgenumber1}
|S| + |S^*| = |E|
\ee
For $N=1$, we have $n(X)=n(Y)=1$, and (\ref{xy}) reduces to
\be
|X|+|Y| = |R|+1
\ee
a relation used in
\cite{WuLu}.  For $N=2$, (\ref{xy}) reduces to Eq. (21) of
Ref. \cite{Ki}.\footnote{After the replacements of $n(X) \to 2-\chi(S),
 n(Y) \to 2-\chi(S^*)$, and $|R|-N \to N-2$.}

\subsection{First identity}
Our construction of correlation duality relations
is based on the use of two identities.
The first identity, which we now state as a lemma,
is a generalization of the identity
(\ref{iden1}).
The new ingredient here is  that the generalization should permit
the consideration of
the factor $v^{|S|} \,\, q^{p(S) - |X|}$
 appearing in the F-K representation (\ref{def:F_X}).

\medskip
\begin{lemma}
For a fixed  edge set $S$ on $G$,  define $X = \pi(S)$ and
$Y = \pi(S^*)$. Then we have the identity
\be\label{thm:iden1}
v^{|S|} \,\, q^{p(S) - [|X| - n(X)]/2} =
k' \,\, (v^{-1} q)^{|S^*|} \,\, q^{p(S^*) - [|Y| - n(Y)]/2}
\ee
where
 \be\label{def:k'}
k' = v^{|E|} \, q^{1 - |F| - [|R| - N]/2}
\ee
\end{lemma}
Remark:
 Notice that $v^{-1} q = e^{K^*} - 1$, so we define
$v^* = v^{-1} q$ as before, and we
can rewrite (\ref{thm:iden1}) as follows to indicate the explicit
dependence on $S$ and $S^*$,
\be\label{iden1'}
v^{|S|} \,\, q^{p(S) - [|\pi(S)| - n(\pi(S))]/2} =
k' \,\, (v^*)^{|S^*|} \,\, q^{p(S^*) - [|\pi(S^*)| - n(\pi(S^*))]/2}
\ee
Proof:  The proof of the lemma parallels  that of the partition function
duality
given in Section 2.2.
First, the relation
(\ref{graphdual}) still holds so that after using (\ref{edgenumber1})
the first line of (\ref{iden1}) becomes
\be\label{iden222}
v^{|S|}\,q^{p(S)}= v^{|E|- |S^*| } q^{|V|-|S| +p(S^{(\rm D)})-1}
\ee
The lemma (\ref{thm:iden1}) now follows after  the use of the
  identities (\ref{y}) and   (\ref{xy}).
Notice that  we have
\be
k' \cdot(k')^* = v^{|E|} \, q^{1 - |F| - [|R| - N]/2}
    \cdot (v^*)^{|E|} \, q^{1 - |V^*| - [|R| - N]/2} =1
\ee
where $|V^*|$, the number of faces of the dual of $G^*$, is  given by
(\ref{Vstar}).

\subsection{Second identity}
Let $S$ and $T$ be edge sets which define the same
partition of $R$, {\it i.e.}, $\pi(S) = \pi(T)$, then a key
feature of  the case
$N=1$  is that their dual edge
sets also define the same partition of $R^*$.   Namely,
\be\label{pidual}
\pi(S) =\pi(T)
{\rm \ \ \ \ if,\>\>and\>\>only\>\>if\ \ \ \  } \pi(S^*) =\pi(T^*), \hskip
1cm N=1
\ee
This feature  permits us to
derive  correlation duality relations for $N=1$  \cite{wu} - \cite{WuLu}.
But (\ref{pidual}) is false for $N \geq 2$  \cite{Ki}.
However a weaker statement holds which we now state as a lemma.

 First, some definitions: A block of $X$ is {\it local} if it contains
roots from one face only,
and  is {\it non-local} otherwise.
A partition $X$ is local if every block of $X$ is local, and is non-local
  if any block in $X$ is non-local.
Given  a partition $X$, we can construct from it a unique local
partition $X_{\rm loc}$ as follows:
We split up every non-local block into a collection of local
blocks, by separating its roots which lie on different faces.
The collection of all these local blocks is the new partition
$X_{\rm loc}$. See Fig. 1 for an example.

Any local partition is a collection of planar partitions
of the roots on each of the external faces.  Each of these planar
partitions on a single face has a unique dual partition. So for
any local partition $\xi = X_{\rm loc}$, we will  define
${\xi}^*$ to be the collection of these dual partitions on $G^*$,
and by extension we will call it the dual partition of $\xi$. Notice that
${\xi}^*$ is always a local partition on $G^*$.

 \medskip
\begin{lemma}
For any edge set $S$ on $G$, we have
\be\label{iden2:1}
\pi(S^*)_{\rm loc} \preceq \pi(S)_{\rm loc}^*
\ee
\medskip
Furthermore, let $S$ and $T$ be two edge sets on $G$, and suppose that
$\pi(S) = \pi(T)$.
Then we have
\be\label{iden2:2}
\pi(S^*)_{\rm loc} = \pi(T^*)_{\rm loc}
\ee
\end{lemma}
Remark: The result corresponding to (\ref{iden2:2}) also holds on $G^*$,
namely, if
$\pi(S^*) = \pi(T^*)$ then also
$\pi(S)_{\rm loc} = \pi(T)_{\rm loc}$.

\bigskip
\noindent Proof:
It is sufficient to establish the relations (\ref{iden2:1}) and
(\ref{iden2:2}) for each external
face separately. So pick one external face, call it $f$, and
let $R_f$ denote the roots which belong to this face, and $R_{f}^*$ their
dual roots.
We will construct two different $N=1$ rooted graphs from $(G,R)$, as follows
(recall that a rooted graph is an {\it ordered pair},
consisting of a graph together with a set of
roots. Mostly we have ignored the distinction between graph and rooted graph,
but we will distinguish them throughout this proof).

First, by ignoring the roots on the other faces we obtain the rooted graph
$(G, R_f)$. As a shorthand we will denote this pair by $G_f$.
Second, by fusing the roots on the other external faces (but not on $f$)
we obtain a different rooted graph $(G^{(F)}_f, R_f)$, which also has roots
only on the face $f$. (Recall that we fuse roots on a face by
merging them into a single vertex).
Again as a shorthand we denote this
by $G^F_f \equiv (G^{(F)}_f, R_f)$.

Each of these rooted graphs has a dual rooted graph,
which we denote by $G_f^*$ and $(G^F_f)^*$ respectively. These are
constructed according to the procedure described in Section 3.1 for the
$N=1$ case, using only the roots on the face $f$. Notice that the
underlying graph of
$(G^F_f)^*$ is identical to the graph $G^*$, because the process of fusing
roots
on a face in $G$ automatically splits the dual vertex in $G^{(\rm D)}$ to
produce the dual roots.
As a convenient shorthand we write $(G^*)_f$ to denote the pair
$(G^*,R_{f}^*)$,
which is obtained from $G^*$ by
ignoring the roots on the other faces. It follows then that
\be\label{iden:graph1}
(G^F_f)^* = (G^*)_{f}
\ee

Now let $S$ be an edge set on $G$. Since the underlying graphs of
$G_f$ and $G^F_f$ have the same
edges as $G$, $S$ also defines partitions of the roots $R_f$ on
$G_f$ and $G^F_f$. For clarity in this proof, we write these partitions as
$\pi(S; G_f)$ and $\pi(S; G^F_f)$ respectively, where we include the
graph in order to distinguish them.
Note that $\pi(S; G_f)$ is one part of the local partition $\pi(S)_{\rm loc}$,
namely the part that contains the roots on $f$. Let $\pi(S)_{f}$
denote this part of the partition $\pi(S)_{\rm loc}$, and
similarly $\pi(S^*)_{f}$ the part of $\pi(S^*)_{\rm loc}$ which contains
the dual roots on $f$. So we have
\be\label{iden:graph2}
\pi(S; G_f) = \pi(S)_{f}, \quad\quad
\pi(S^*; (G^*)_{f}) = \pi(S^*)_{f}
\ee

To prove the first result (\ref{iden2:1}), note that
since $G^F_f$ is obtained from $G_f$ by fusing vertices we have
\be
\pi(S; G_f) \preceq \pi(S; G^F_f)
\ee
These are both $N=1$ partitions, hence they have dual partitions,
and these satisfy
\be
\pi(S; G_f)^* \succeq \pi(S; G^F_f)^* = \pi(S^*; (G^F_f)^*)
\ee
Using (\ref{iden:graph1}) and (\ref{iden:graph2}) this gives
\be
\pi(S)_{f}^* = \pi(S; G_f)^* \succeq \pi(S^*; (G^*)_{f})
\ee
Combining these for every external face gives
\be
\pi(S)_{\rm loc}^* \succeq \pi(S^*)_{\rm loc}
\ee

To prove the second result (\ref{iden2:2}), note that since
$\pi(S) = \pi(T)$ it follows also that
\be
\pi(S; G^F_f) = \pi(T; G^F_f).
\ee
Taking the dual of both sides gives
\be
\pi(S^*; (G^F_f)^*) = \pi(T^*; (G^F_f)^*)
\ee
and hence
\be
\pi(S^*; (G^*)_{f}) = \pi(T^*; (G^*)_{f})
\ee
Again combining these for every external face gives
\be
\pi(S^*)_{\rm loc} = \pi(T^*)_{\rm loc}
\ee
Q.E.D.

\subsection{Form of duality relations}
As a convenient shorthand, for  partitions $X$ of $R$ and $Y$ of $R^*$ we write
\be
Z_X = Z_X(G;q,v), \hskip 1cm
Z_{Y}^* = Z_Y(G^*;q,v^*)
\ee
and similarly we write
\be
F_X = F_X(G;q,v), \hskip 1cm
F_{Y}^* = F_Y(G^*;q,v^*)
\ee
where $F_X$ is defined in (\ref{F-Kcorr}).
We seek duality relations which
express each given $Z_X$  in terms of a linear combination of the
dual $Z_Y^*$.
Recall the M\"obius inversion
(\ref{Mobius}) relating $Z_X$ to $F_X$,
it is  sufficient to obtain duality relations for
the functions $F_X$.  Indeed,
   for $N=1$, for example,
  the identity (\ref{pidual}) ensures that there is
a one-to-one correspondence between the (connected) partitions $X=\pi(S)$
on $G$ and $Y=\pi(S^*)$ on $G^*$.  Then, using Lemma 1 one obtains the
desired duality relation
\be\label{Nonedual}
q^{|X|/2} F_X(G;q,v) = k'\ q^{|Y|/2} F_Y(G^*;q,v^*) \hskip 1cm N=1
 \ee
where we have used  $n(X) = n(Y) =1$, and the coefficient $k'$ was defined
in (\ref{def:k'}).  This duplicates a result of \cite{WuLu}.\footnote{It
can be
readily verified that Eq. (\ref{Nonedual}) is the same as Eq. (49) of
\cite{WuLu}, after the substitution of $F_X=D_X$ and $F^*_Y=q^{-|E|/2} D^*_Y$.}

For $N\geq 2$, however, the mappings $X \to Y$
and/or $Y\to X$ are  not necessarily one-to-one \cite{Ki}.  See Fig. 2
for an example.  Generally,
for fixed $\pi(S)=X,\pi(S^*)=Y$, we define by analogy to (\ref{def:F_X})
the functions
\bea\label{def:F_XY}
F_{X,Y}(G;q,v) &=& \sum_{S \in {\cal S}(X),S^*\in {\cal S}(Y)}
v^{|S|} \,\, q^{p(S) - |X|} \nonumber \\
F_{Y,X}(G^*;q,v^*) &=& \sum_{S \in {\cal S}(X),S^*\in {\cal S}(Y)}
{(v^*)}^{|S^*|} \,\, q^{p(S^*) - |Y|},
\eea
Then, one has the duality relation
\be\label{Ndual}
q^{|X|/2} F_{X,Y}(G;q,v) = k'\ q^{|Y|/2} F_{Y,X}(G^*;q,v^*)
 \ee
which is a generalization of (\ref{Nonedual}).
The partial partition functions are then
\bea\label{partialXY}
Z_X(G;q,v)&=& \sum_{X'\preceq X} \sum_Y F_{X,Y}(G;q,v) \nonumber  \\
Z_Y(G^*;q,v^*)&=& \sum_{Y'\preceq Y} \sum_X F_{Y,X}(G^*;q,v^*)
\eea
For $N=1$, the $X\leftrightarrow Y$ mapping is unique,
 so that (\ref{partialXY})
becomes (\ref{F-Kcorr}) and the relation (\ref{partialXY}) can be inverted,
permitting one to express $Z_X$ as a linear combination of $Z_Y$'s.
For $N\geq 2$, however, (\ref{partialXY}) cannot be inverted,
and so we cannot express the $F_{X,Y}$ in terms of the $Z_X$. Hence the duality
relations (\ref{Ndual}) do not help us.
As a result, we
  will instead derive
duality relations of the form
\be\label{def:dualF}
\sum_{X} \alpha (X) F_X = k' \, \sum_{Y} \beta (Y) F_{Y}^*
\ee
The sum on the left runs over partitions of the root set $R$,
and the sum on the right side runs over partitions on $R^*$.
 The coefficients $\alpha$ and $\beta$ depend on
$q,v$ as well as their arguments $X$ and $Y$, and are different for
each duality relation.

Using the M\"obius inversion (\ref{Mobius}), (\ref{def:dualF}) provides
duality relations for the partial partition functions, and we end up with
\be\label{def:dualZ}
\sum_{W} a (W) Z_W = k' \, \sum_{U} b (U) Z_{U}^*,
\ee
where
\bea
a (W) &=& \sum_{X \succeq W} \mu (W,X) \alpha (X) \nonumber \\
b (U) &=& \sum_{Y \succeq U} \mu (U,Y) \beta (Y)
\eea
Note that the number of independent relations depends on
the root set $R$. If all roots lie on one face, so that $N=1$,
 then our result produces the same number of relations as
the number of planar partitions of the root set. This duplicates
the results of \cite{WuLu}.
If all roots lie on distinct faces, so
that $|R| = N$, then our method produces just one relation, namely
the original duality (\ref{origdual}).

\subsection{Correlation duality}
We use the two lemmas to generate correlation duality relations.
The definition (\ref{def:F_XY}) was not useful because it involved a
sum over all edge sets $S$ with fixed partition $X=\pi(S)$ and fixed
dual partition $Y=\pi(S^*)$. As we now show, one way to
generate duality relations is by summing instead over all edge
sets with {\it fixed local} partitions $X_{\rm loc}$ and $Y_{\rm loc}$.
According to (\ref{iden2:1}) each of these must be
a refinement of the dual of the other, so we can restrict attention
to such pairs of local partitions.

Accordingly, writing 
\be
\xi = X_{\rm loc}, \hskip 1cm \eta = Y_{\rm loc}
\ee
we say that $(\xi, \eta)$ are {\it compatible} if
\be\label{compat}
\eta \preceq {\xi}^* \quad {\rm and} \quad \xi \preceq {\eta}^*
\ee

Let $(\xi, \eta)$ be compatible local partitions, and
define a collection of edge sets on $G$ as follows:
\be\label{def:calS2}
{\cal S}(\xi, \eta) = \{S \subset E \,:\, \pi(S)_{\rm loc} = \xi, \,\,
\pi(S^*)_{\rm loc} = \eta \}
\ee
For ease of notation, we write ${\cal S}(\xi, \eta)^*$ for the collection of
edge sets on $G^*$ which are dual to those in ${\cal S}(\xi, \eta)$.

In general, for a given pair $(\xi,\eta)$, the set
${\cal S}(\xi, \eta)$ may be empty. (For example, in the case $N=1$ this
happens
unless $\eta = {\xi}^*$). However if ${\cal S}(\xi, \eta)$ is not empty,
then (\ref{iden2:2}) implies that it must have the following form
(recall the definition (\ref{def:calS1}))
\be\label{union}
{\cal S}(\xi, \eta) = {\cal S}(X_1) \cup \cdots \cup {\cal S}(X_m)
\ee
for {\it some} collection of partitions $\{X_1, \dots, X_m \}$.
The reason is clear: suppose that $S \in {\cal S}(\xi, \eta)$ and that
$\pi(S) = X$. Let $T$ be any other edge set with $\pi(T) = X$. Then by
definition
also $\pi(T)_{\rm loc} = \xi$, and by (\ref{iden2:2}),
$\pi(T^*)_{\rm loc} = \eta$. Hence also $T \in {\cal S}(\xi, \eta)$, hence
${\cal S}(X) \subset {\cal S}(\xi, \eta)$.

We postpone for a moment the question of determining {\it which}
partitions occur on the right side of
(\ref{union}). First we will use this relation to write the sum over edge sets
in ${\cal S}(\xi, \eta)$ as a sum over factors $F_{X}$ defined in
(\ref{def:F_X}). Recalling the left side of
(\ref{iden1'}), and using (\ref{union}) and (\ref{def:F_X}) we get the
following identity:
\bea\label{eqn1}
\sum_{S \in {\cal S}(\xi, \eta)} v^{|S|} \,\, q^{p(S) -  [|\pi(S)| -
n(\pi(S))]/2} & =  &
\sum_{j=1}^{m} \sum_{S \in {\cal S}(X_j)}
v^{|S|} \,\, q^{p(S) - [|\pi(S)| - n(\pi(S))]/2} \nonumber \\
& = &
\sum_{j=1}^{m} q^{[|X_j| + n(X_j)]/2} \,\, F_{X_j} \nonumber \\
\eea

Next we use our first identity.
Using (\ref{iden1'}) we can rewrite the left side of (\ref{eqn1}) as
\be\label{eqn2}
\sum_{S \in {\cal S}(\xi, \eta)} v^{|S|} \, q^{p(S) - [|\pi(S)| -
n(\pi(S))]/2} =
k'  \sum_{S^* \in {\cal S}(\xi, \eta)^*}
(v^*)^{|S^*|} \, q^{p(S^*) -[|\pi(S^*)| - n(\pi(S^*))]/2}
\ee
Now we repeat for $G^*$ the argument leading to (\ref{union}),
and  obtain
\be\label{union*}
{\cal S}(\xi, \eta)^* = {\cal S}(Y_1) \cup \cdots \cup {\cal S}(Y_l)
\ee
for some collection of partitions $\{Y_1, \dots, Y_l \}$ of $R^*$.
Hence we end up with the analog of (\ref{eqn1}), namely
\be\label{eqn1*}
\sum_{S^* \in {\cal S}(\xi, \eta)^*}
(v^*)^{|S^*|} \,\, q^{p(S^*) - [|\pi(S^*)| - n(\pi(S^*))]/2} =
\sum_{i=1}^{l} q^{[|Y_i| + n(Y_i)]/2} \,\, F_{Y_i}^*
\ee

Putting together (\ref{eqn1}), (\ref{eqn2}) and (\ref{eqn1*}) we
get
\be\label{eqn3}
\sum_{j=1}^{m} q^{[|X_j| + n(X_j)]/2} \,\, F_{X_j}
= k' \,\, \sum_{i=1}^{l} q^{[|Y_i| + n(Y_i)]/2} \,\, F_{Y_i}^*
\ee
This is our new duality relation. As stated earlier, it relates a linear
combination of $F_X$ to a linear combination of $F_{Y}^*$.

Now we turn to the question of which partitions can occur on the left-hand
and right-hand sides of (\ref{eqn3}). These are determined by the compatible pair
$(\xi, \eta)$ via the relations (\ref{union}) and (\ref{union*}).
In order to decide which partitions occur in (\ref{union}) and
(\ref{union*}),
it seems to be necessary to
work on a case by case basis, by starting with
$(\xi, \eta)$ and
examining each planar partition $X$ to see if it produces this pair.
So partly to hide our ignorance, and partly to
tidy up the notation, we define
\be
\theta(X;\xi,\eta) = \cases{1 & if $\exists$ $S \subset E$ such that
$\pi(S)=X$, $\pi(S)_{\rm loc}=\xi$, $\pi(S^*)_{\rm loc}=\eta$\cr
0 & otherwise}
\ee
Similarly if $Y$ is a partition of $R^*$, then define
${\theta}^*(Y;\xi,\eta) =1$
if there is an edge set $S^*$ with $\pi(S^*)=Y$ and also
$\pi(S)_{\rm loc}=\xi$, $\pi(S^*)_{\rm loc}=\eta$, and otherwise
${\theta}^*(Y;\xi,\eta) =0$.
Then we can rewrite (\ref{eqn3}) as
\bea\label{eqn4}
\sum_{X} \theta(X;\xi,\eta) q^{[|X| + n(X)]/2}  \,\, F_{X} & & \nonumber \\
 =  k' \,\, \sum_{Y} {\theta}^*(Y;\xi,\eta) &
q^{[|Y| + n(Y)]/2} \,\, F_{Y}^*,
\eea
and this is precisely the form described in (\ref{def:dualF}).

Finally we use the M\"obius relation (\ref{Mobius}) to
re-express (\ref{eqn4}) in terms of the partial partition functions.
The result has the form (\ref{def:dualZ}), namely
\be\label{ab}
\sum_{W} a (W) Z_W = k' \, \sum_{U} b (U) Z_{U}^*,
\ee
where the coefficients are
\be
a(W) = a(W;\xi,\eta) =
\sum_{X \succeq W} \,\mu(W,X) \,\theta(X;\xi,\eta) \, q^{[|X| + n(X)]/2}
\ee
and
\be
b(U) = b(U;\xi,\eta) =
\sum_{Y \succeq U} \,\mu(U,Y) \,{\theta}^*(Y;\eta,\xi) \, q^{[|Y| + n(Y)]/2}
\ee
To summarize,  we have obtained a duality relation for every pair of
compatible local partitions $(\xi, \eta)$ in the
form of (\ref{eqn4}). In some cases the relation is empty;
otherwise it is given by (\ref{eqn4}).  However, it must be emphasized that
while we have obtained new duality relations for the partial partition
functions
$Z_X$, our prescription does not necessarily generate {\it all} duality relations
(in fact we know that it is incomplete in the case $N=2$ as described below).
The generation of the complete set of dualities remains an open question.

\section{Examples}
\subsection{Two external faces, two roots each}
This case was examined in detail in \cite{Ki}
(see in particular Figs. 6 and 7), and we compare our results
here with those in \cite{Ki}. Label the roots $1,2$ on one face,
and $3,4$ on the other. There are four local partitions on $G$, namely
$(1)(2)(3)(4)$, $(12)(3)(4)$, $(1)(2)(34)$ and $(12)(34)$
(we follow standard notation for partitions, so, for example,
$(1)(2)(34)$ means that there are
three blocks, containing roots $(1)$, $(2)$ and $(3,4)$ respectively). Since
there are
also four local partitions on $G^*$, there are 16 possible pairs
$(\xi, \eta)$. However, only 10 of these satisfy (\ref{iden2:1}),
so there are 10 possible duality relations (\ref{ab}).
Closer examination shows that only 5 of these pairs can occur
as local partitions of edge sets. We write
$1^*, 2^*, 3^*, 4^*$ for the dual roots.
The 5 possible pairs are
\bea\label{5 pairs}
\xi = (1)(2)(3)(4), \quad \eta & = & (1^{*}2^{*})(3^{*}4^{*}) \\ \nonumber
\xi = (1)(2)(34), \quad \eta & = & (1^{*}2^{*})(3^{*})(4^{*}) \\ \nonumber
\xi = (12)(3)(4), \quad \eta & = & (1^{*})(2^{*})(3^{*}4^{*}) \\ \nonumber
\xi = (12)(34), \quad \eta & = & (1^{*})(2^{*})(3^{*})(4^{*}) \\ \nonumber
\xi = (1)(2)(3)(4), \quad \eta & = & (1^{*})(2^{*})(3^{*})(4^{*})
\eea

There are 15 partitions of the 4 roots on $G$, and they are all planar,
namely, each one can occur as $\pi(S)$ for some edge set $S$.
By our basic result (\ref{union}) each partition is associated
with one of the five pairs in (\ref{5 pairs}). Similarly the 15 partitions
of the dual roots are each associated with one pair.
In Table 1 we show these associations.

The 5 duality relations are now obtained by substituting into (\ref{eqn3}).
For example, the third pair in (\ref{5 pairs}) gives the identity
\bea\label{third pair}
& q^{5/2} & F_{(12)(3)(4)}  + q^{3/2} F_{(123)(4)} + q^{3/2} F_{(124)(3)}
\\ \nonumber
& = &
k' [q^{5/2} F_{(1^{*})(2^{*})(3^{*}4^{*})} +
q^{3/2} F_{(1^{*}3^{*}4^{*})(2^{*})} +
q^{3/2} F_{(2^{*}3^{*}4^{*})(1^{*})}]
\eea
This reproduces a result in \cite{Ki}.\footnote{It
can be verified that Eq. (\ref{third pair}) is the same as Eq. (29) of
\cite{Ki}, after the substitutions
$T_X = v^{-|E|/2} q^{|F|/2 + |X|} F_X$, $T^{*}_X =
v^{|E|/2} q^{-|F|/2 + |Y|} F_{Y}^{*}$, and renaming roots
$1 \rightarrow 1, 2 \rightarrow 3, 3 \rightarrow 4, 4 \rightarrow 2$,
and dual roots
$5 \rightarrow 1^{*}, 6 \rightarrow 3^{*}, 7 \rightarrow 4^{*},
8 \rightarrow 2^{*}$.
}
Similarly the first, second and fourth pairs in (\ref{5 pairs})
reproduce three other identities in \cite {Ki}.
However the last pair in (\ref{5 pairs}) produces an identity
which is the {\it sum} of two independent identities in \cite{Ki}.

\subsection{The general case $N = 2$}
The general case $N=2$ is similar to the example above. 
The present method reproduces many but not all of the duality identities discovered in
\cite{Ki}. In  \cite{Ki} a non-local block in a partition $X$
was called a {\it bridge}, and $X$ was called a {\it $k$-bridge partition}
if it contained $k$ bridges ($k=0,1,\dots$).
It was found that there exists one independent duality relation corresponding to
every $0$-bridge partition, and in addition one independent relation for every
$k$-bridge partition with $k \geq 2$.

Our method here reproduces all the duality relations for the $0$-bridge
partitions, but {\it not} for the $k$-bridge partitions for $k\geq 2$. In the latter case
it combines $k$ independent relations into a single relation. This was
the case in the previous example, where the two relations for $k=2$ were
combined into a single relation.

\subsection{An example with $N=3$}
As an illustration of our method we consider
the case where there are three external faces,
each containing two roots. Roots $1,2$ are on face 1,
roots $3,4$ are on face 2 and roots $5,6$ are on face 3.
The dual roots are $1^{*}, 2^{*}$ etc.

The number of local partitions
of $R$ is 8, since there are two choices on each face. Hence the total
number of pairs $(\xi, \eta)$ of local partitions on $R$ and $R^*$ is
64. The number of compatible local pairs is 27. After closer analysis,
it turns out that only 15 of these pairs can be realised via edge sets.
Hence our method produces 15 independent duality relations corresponding to
these 15 allowed pairs.

For brevity we present just one of these relations,
corresponding to the following local pair:
\be\label{one pair}
\xi = (12)(3)(4)(5)(6), \quad
\eta = (1^{*})(2^{*})(3^{*}4^{*})(5^{*}6^{*})
\ee
Instead of writing out the identity, we list in Table 2  the 13
partitions which occur on the left-hand side of (\ref{eqn3}),
and the 9 dual partitions that occur on the right-hand side,
along with the exponents of $q$.

\subsection{Bounds on the number of relations}
In general it seems to be a hard problem to determine exactly the number of
compatible local partitions which can produce
duality relations. However our method does provide a lower
bound on this number. If $\xi$ is a local partition of $R$, then
certainly ${\cal S}(\xi, {\xi}^*)$ is not empty, and hence there is always
a duality
relation for this compatible pair. The number of such relations is the
product of
the number of planar local partitions on the faces. For example, if $N
= 2|R|$ and the roots are paired on the faces, then the number of
relations is at least
$2^{N}$.

\section{The Tutte polynomial}
In 1955 Tutte \cite{tutte1,tutte2} introduced in graph theory the
notion of  dichromatic, or Tutte, polynomials, which
turns out to be precisely the Potts partition function.
For our purposes and to conform with notations of
\cite{wukinglu}, we define the Tutte
polynomial associated with a graph $G$ as the two-variable polynomials
\bea\label{tutte}
Q(G;t,v) &=& v^{-|V|}\  Z(G; vt, v) \nonumber \\
    &=& v^{-|V|}\sum_{S \subseteq E} t^{p(S)}v^{|S|+p(S)}
\eea
where we have $t=v^*=q/v$.
Similarly, one defines as in \cite{wukinglu} the rooted Tutte polynomials
 \bea\label{rootedtutteXY}
Q_X(G;t,v) &=& v^{-|V|}\  Z_X(G; vt, v) \nonumber \\
Q_{X,Y}(G;t,v) &=& v^{-|V|}\  Z_{X,Y}(G; vt, v)
\eea
and the associated summations
\bea
 H_X(G;t,v)  &=& v^{-|V|} F_X(G;vt,v) \nonumber \\
 H_Y(G^*;v,t)  &=& t^{-|V^*|} F_Y(G;vt,t) \nonumber \\
 H_{X,Y}(G;t,v)  &=& v^{-|V|} F_{X,Y}(G;vt,v) \nonumber \\
 H_{Y,X}(G^*;v,t)  &=& t^{-|V^*|} F_{Y,X} (G;vt,t)
\eea
Then,  the duality relation (\ref{origdual}) for the partition function
 becomes
\be
v\ Q(G;t,v) = t\ Q(G^{\rm (D)};v,t)
\ee
 and Lemma 1 assumes the form
\be
v^{1-|V|+|X|-n(Y) } \Big[v^{|S|}(vt)^{p(S) -|X|}\Big] =
t^{1-|V^*|+|Y|-n(X)} \Big[t^{|S^*|}(vt)^{p(S^*) -|Y|}\Big]
\ee
 Likewise, the $N=1$ correlation duality (\ref{Nonedual}) becomes
 \be
v^{|X|}\ H_X(G;t,v) = t^{|Y|}\ H_Y(G^*; v,t) \hskip 1cm N=1
\ee
and the duality (\ref{Ndual}) becomes
 \be
v^{|X|}t^{n(X)-1}\ H_{X,Y}(G;t,v) = t^{|Y|}v^{n(Y)-1}\ H_{Y,X}(G^*; v,t)
\ee
Furthermore, the relation (\ref{eqn3}) for general $N$ can be written as
\bea\label{eqn31}
&&v^{1-|V| +(|R|-N)/2}
\sum_{j=1}^{m} (vt)^{[|Y_i| + n(Y_i)]/2} \,\, H_{X_j} \nonumber \\
&&\hskip 3cm = t^{1-|V^*| +(|R|-N)/2}
 \,\, \sum_{i=1}^{l} (vt)^{[|Y_i| + n(Y_i)]/2} \,\, H_{Y_i}^*
\eea
These expressions reflect  the symmetric roles played
by the variables $v$ and $t=v^*=q/v$.

\bigskip
\noindent
{\bf Acknowledgments}\par
This research was supported in part by a Research
Scholarship Development Fund (RSDF) grant from Northeastern
University. In addition CK was
partially supported by NSF Grant DMS-9705779, and
FYW was partially supported by NSF Grant DMR-9980440.
We thank W. T. Lu for help in preparing the figures.

\vfill\eject

\begin{table}
\centering
\begin{tabular}{|l|l|l|} \hline
Local pair $(\xi, \eta)$ & Partitions of $R$ &
Partitions of $R^{*}$ \\ \hline
(1)(2)(3)(4)                & (1)(2)(3)(4) & $(1^{*}2^{*})(3^{*}4^{*})$ \\
$(1^{*}2^{*})(3^{*}4^{*})$    & (13)(2)(4)  & $(1^{*}2^{*}3^{*}4^{*})$ \\
                            & (14)(2)(3) & \\
                            & (23)(1)(4) & \\
                            & (24)(1)(3) &  \\ \hline
(1)(2)(34)                  & (1)(2)(34) & $(1^{*}2^{*})(3^{*})(4^{*})$\\
$(1^{*}2^{*})(3^{*})(4^{*})$  & (134)(2) & $(1^{*}2^{*}3^{*})(4^{*})$\\
                            & (234)(1) & $(1^{*}2^{*}4^{*})(3^{*})$\\ \hline
(12)(3)(4)                  & (12)(3)(4) & $(1^{*})(2^{*})(3^{*}4^{*})$\\
$(1^{*})(2^{*})(3^{*}4^{*})$  & (123)(4) & $(1^{*}3^{*}4^{*})(2^{*})$\\
                            & (124)(3) & $(2^{*}3^{*}4^{*})(1^{*})$\\ \hline
(12)(34)                    & (12)(34) & $(1^{*})(2^{*})(3^{*})(4^{*})$\\
$(1^{*})(2^{*})(3^{*})(4^{*})$  & (1234) & $(1^{*}3^{*})(2^{*})(4^{*})$\\
                            &  & $(1^{*}4^{*})(2^{*})(3^{*})$\\
                            &  & $(2^{*}3^{*})(1^{*})(4^{*})$\\
                            &  & $(2^{*}4^{*})(1^{*})(3^{*})$\\ \hline
(1)(2)(3)(4)                & (13)(24) & $(1^{*}3^{*})(2^{*}4^{*})$\\
$(1^{*})(2^{*})(3^{*})(4^{*})$  & (14)(23) & $(1^{*}4^{*})(2^{*}3^{*})$\\
\hline
\end{tabular}
\caption{List of compatible local pairs and their associated partitions.}
\end{table}

\clearpage

\begin{table}
\centering
\begin{tabular}{|l|l|l|l|} \hline
Partition $X$ & $|X| + n(X)$ & Partition $Y$ &
$|Y| + n(Y)$ \\ \hline
(12)(3)(4)(5)(6)    & 8 & $(1^{*})(2^{*})(3^{*}4^{*})(5^{*}6^{*})$ & 7 \\
(123)(4)(5)(6)    & 6 & $(1^{*}3^{*}4^{*})(2^{*})(5^{*}6^{*})$ & 5 \\
(124)(3)(5)(6)    & 6 & $(1^{*}5^{*}6^{*})(2^{*})(3^{*}4^{*})$ & 5 \\
(125)(3)(4)(6)    & 6 & $(1^{*})(2^{*}3^{*}4^{*})(5^{*}6^{*})$ & 5 \\
(126)(3)(4)(5)    & 6 & $(1^{*})(2^{*}5^{*}6^{*})(3^{*}4^{*})$ & 5 \\
(12)(35)(4)(6)    & 6 & $(2^{*})(1^{*}3^{*}4^{*}5^{*}6^{*})$ & 3 \\
(12)(36)(4)(5)    & 6 & $(1^{*})(2^{*}3^{*}4^{*}5^{*}6^{*})$ & 3 \\
(12)(3)(45)(6)    & 6 & $(1^{*}3^{*}4^{*})(2^{*}5^{*}6^{*})$ & 3 \\
(12)(3)(46)(5)    & 6 & $(2^{*}3^{*}4^{*})(1^{*}5^{*}6^{*})$ & 3 \\
(1235)(4)(6)    & 4 &  &  \\
(1236)(4)(5)    & 4 &  &  \\
(1245)(3)(6)    & 4 &  &  \\
(1246)(3)(5)    & 4 &  &  \\ \hline
\end{tabular}
\caption{List of partitions and dual partitions for the pair (\ref{one pair}).}
\end{table}

\clearpage
\vskip 2cm

\centerline{\bf Figure Captions}

\bigskip
\noindent
Fig. 1. An $N=3$ example of a partition
$X$  and the associated local partition $X_{\rm loc}$.
(a) $X=(135)(247)(68)$,
(b) $X_{\rm loc} =(1)(35)(2)(4)(7)(68)$.
Cross-hatched areas denote external faces; lines denote connected edge sets.

\bigskip
\noindent
Fig. 2. An $N=2$ example showing a given  partition $X =(1)(23)(4)$ (solid line)
can be the dual of two different partitions $Y_1$ and $Y_2$ (broken lines).
\newline (a) $Y_1 =(1^*2^*3^*4^*)$,
(b)   $Y_2=(1^*2^*)(3^*4^*)$.

\vfill\eject

\begin{figure}
\centering{\resizebox{!}{2.5in}{\includegraphics{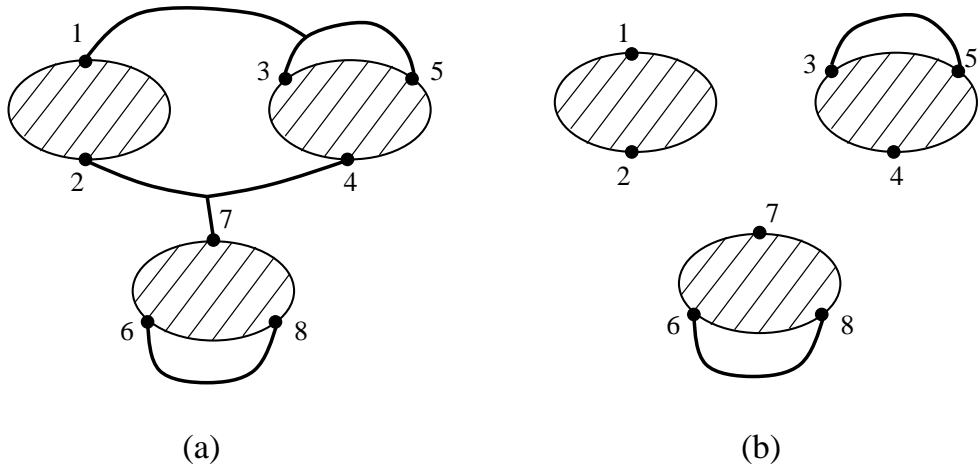}}}
\caption{An $N=3$ example of a partition
$X$  and the associated local partition $X_{\rm loc}$.
(a) $X=(135)(247)(68)$,
(b) $X_{\rm loc} =(1)(35)(2)(4)(7)(68)$.
Cross-hatched areas denote external faces; lines denote connected edge sets.
\label{fig1}}
\end{figure}

\clearpage

\begin{figure}
\centering{\resizebox{!}{2.0in}{\includegraphics{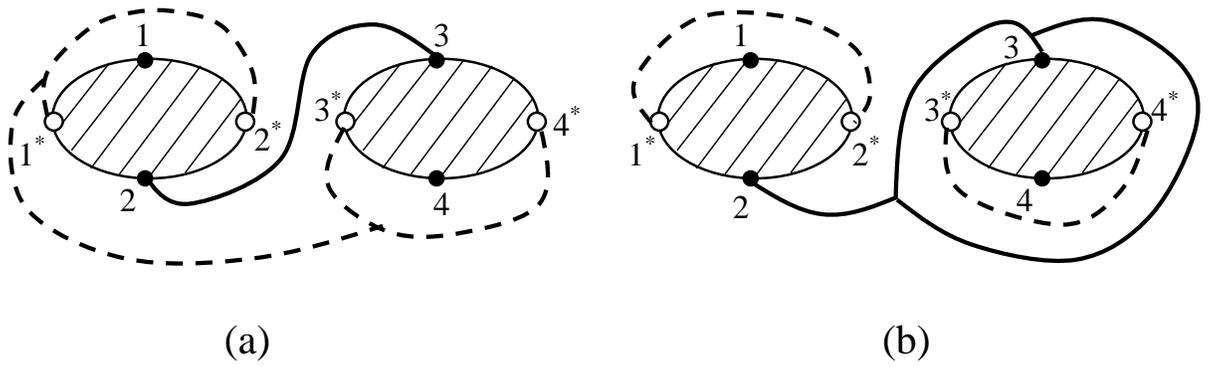}}}
\caption{An $N=2$ example showing a given  partition $X =(1)(23)(4)$ (solid line)
can be the dual of two different partitions $Y_1$ and $Y_2$ (broken lines).
\newline (a) $Y_1 =(1^*2^*3^*4^*)$,
(b)   $Y_2=(1^*2^*)(3^*4^*)$.
\label{fig2}}
\end{figure}

\end{document}